**Equivalent slip length of flow around a super-hydrophobic cylinder**

Zhi-yong Li (李志勇)[1], Ya-kang Xiao (肖雅康)[2], Yan-cheng Li (李晏丞)[3],

Li Yu (于丽)[4], Sai Peng (彭赛)[2,5]*, Yong-liang Xiong (熊永亮)[6]*

[1]Department of Energy and Power Engineering, School of Chemical Engineering and Energy Technology, Dongguan University of Technology, China

[2]National Center for Applied Mathematics in Hunan, Xiangtan, 411105, China

[3]School of Naval Engineering, Wuxi Institute of Communications Technology, 214151, Wuxi, China

[4]School of Civil Engineering and Architecture, Southwest University of Science and Technology, Mianyang, 621010, China

[5]School of Mathematics and Computational Science, Xiangtan University, Xiangtan, 411105, China

[6]Department of Mechanics, Huazhong University of Science and Technology, Wuhan 430074, China

**Abstract**

In this research, a two-dimensional numerical simulation is conducted to determine the equivalent wall slip length for flow around a circular cylinder featuring a super-hydrophobic surface. The super-hydrophobic surface is modeled as an alternating distribution of slip and no-slip conditions along the cylinder's surface. The smallest unit of this alternating pattern is referred to as a monomer. The study takes into account the Reynolds number and two critical dimensionless parameters: the gas fraction (*GF*) and the ratio *l*/*a*. *GF* indicates the proportion of the slip length relative to the total length of the monomer, while *l*/*a* denotes the ratio of the monomer length (*l*) to the cylinder's radius (*a*). The ranges considered for the Reynolds number, *GF*, and *l*/*a* are from 0.2 to 180, 0.1 to 0.99, and π/80 to π/5, respectively. A dimensionless number, the Knudsen number (*Kn*), is introduced to measure the ratio between the equivalent slip length and the cylinder's diameter (*D*). By equating the integral wall friction resistance on the cylinder surface, a quantitative relationship between the equivalent *Kn* and the parameters (*Re*, *GF*, *l*/*a*) is established. A meticulous comparison of flow parameters between the equivalent slip length model and the slip-no-slip scenario reveals that the slip length model is an effective approximation

Corresponding Author: Sai Peng (pscfd@xtu.edu.cn); Yong-liang Xiong (xylcfd@outlook.com)

for the slip-no-slip alternating model.

## 1. Introduction

The wall no-slip hypothesis, a fundamental tenet in fluid mechanics, posits that the fluid velocity at the wall is equal to the wall's velocity, typically zero for a stationary surface. However, numerous recent experiments[1,2] have challenged this assumption, revealing its potential inaccuracy under specific conditions, particularly in aqueous environments. For example, air may become trapped in the microstructures of a superhydrophobic surface, preventing direct contact between the water and the wall and resulting in a slip velocity at the wall. [3-5] Investigating wall slip phenomena is not only of significant theoretical importance for comprehending flow behavior under slip conditions but also holds potential for engineering applications. Wall slip could potentially be used to reduce wall frictional resistance, providing substantial benefits across various industrial and environmental scenarios.[6] In this study, the primary focus is on the phenomenon of wall slip as it pertains to superhydrophobic surfaces in aqueous environments.

Flow past a cylinder is a central topic within the fields of fluid dynamics and hydraulic engineering, representing one of the most intricate challenges among various flow patterns. The simplest scenario involves viscous flow past a confined circular cylinder, where the flow pattern evolves through multiple stages as the Reynolds number (*Re*) changes.[7-9] Indeed, the presence of wall slip has a profound effect on the flow dynamics surrounding a circular cylinder. Studies, both experimental[10-13] and numerical[14-21], have demonstrated that wall slip can significantly alter the flow characteristics. It can reduce the drag force on the cylinder, diminish flow fluctuations, shorten the wake region, and enhance the frequency of vortex shedding, among other effects. These alterations to the flow field highlight the critical need to account for wall slip when analyzing and designing fluid flow systems, emphasizing its role in influencing overall flow performance and efficiency.

Firstly, we review the experimental research on the flow over a super-hydrophobic circular cylinder. Muralidhar *et al.*[10] investigated the impact of partial slip conditions on different superhydrophobic surfaces at high Reynolds numbers, up to 10,000. They observed that flow behavior is contingent upon the arrangement of ridges on these surfaces. The shedding frequency was found to be higher compared to a smooth cylinder, with ridges aligned in the flow direction exhibiting a more pronounced increase in shedding frequency than those arranged normal to the flow. They also noted a delayed initiation of vortex shedding and an elongation of the recirculation bubble on a superhydrophobic cylinder surface. Kim & Park[12] explored the impact of slip direction on the flow field and drag coefficient by sanding hydrophobic cylinder surfaces with sandpapers of two different grit sizes in streamwise and spanwise directions. Their results indicated that a rough hydrophobic surface intensifies turbulence in the flow above the circular cylinder and along the separating shear layers. This leads to a delay in flow separation, earlier vortex roll-up in the wake, and a reduction in the size of the recirculation bubble in the wake by up to 40%. Meanwhile, a drag reduction of less than 10% was estimated from wake surveys for Reynolds numbers between 0.7 and $2.4\times10^4$. Those two experimental studies suggest that the orientation of superhydrophobic structures affects the flow velocity slip behavior on the cylinder wall. Brennan *et al.*, [13] by applying superhydrophobic sand of sizes ranging from 50 to 710 μm to increase surface roughness, achieved a maximum drag reduction of 28% within the *Re* range of 10,000 to 40,000. The experimental study implies that the arrangement of superhydrophobic structures also affects the flow velocity slip behavior on the cylinder wall.

Secondly, we review the numerical research and theoretical studies. Currently, there are two predominant approaches to simulate the flow around superhydrophobic cylinders. Initially, the Navier-slip condition,[22] characterized by the slip length, is employed to depict the superhydrophobic wall behavior. Employing this methodology, an extensive two-dimensional direct numerical simulation (2D-DNS) was conducted by Legendre *et al.*[14] to identify the critical Knudsen number above which the downstream recirculation bubbles and wake vortex shedding in the flow around a circular cylinder vanish. Seo *et al.* [15] conducted a numerical simulation of shallow-water flow over a cylinder to examine the flow parameters, and discovered that wall slip mitigated the wall vortex. Additionally, the shear layer developing along the cylinder surface, influenced by the Navier-slip condition, tended to shift the separation point rearward towards the stagnation point. Li *et al.*[16] conducted a numerical investigation of two-dimensional flow past a confined circular cylinder with a slip wall. In their research, the time-averaged velocity distribution on the cylinder's surface was found to align closely with a specific formula $\overline{u_\tau} = \left[ \frac{\alpha}{1+\beta e^{-\gamma(\pi-\theta)}} + \delta \right] \cdot \sin\theta$, where the coefficients ($\alpha$, $\beta$, $\gamma$, $\delta$) are functions of both the Reynolds number (*Re*) and the Knudsen number (*Kn*). They identified several scaling laws: log($\overline{u_{\tau max}}$)~log(*Re*) and $\overline{u_{\tau max}}$ ~*Kn* for low *Kn* ($\overline{u_{\tau max}}$ is the maximum velocity on the cylinder surface), log(*DR*)~log(*Re*) ($Re \leq 45$ and $Kn \leq 0.1$), log(*DR*)~log(*Kn*) ($Kn \leq 0.05$). Here, *DR* represents the drag reduction ratio. Theoretically, Li *et al.*[17] provide an analytical solution derived from the matched-asymptotic expansion method. This analytical solution is specifically tailored for scenarios with low Reynolds numbers ($Re \ll 1$). In contrast, Kumar *et al.*[18] present an asymptotic theory that addresses the high-Reynolds-number flow past a shear-free circular cylinder, as *Re* approaches ∞.

On the other hand, a slip-no-slip alternating structure is implemented to describe the superhydrophobic behavior of the wall. Li *et al.* [17] investigated the slip distribution's impact on flow over a circular cylinder and determined that the flank slip configuration yielded the most significant drag reduction compared to fore-side slip and aft-slip configurations for Reynolds numbers between 100 and 500 and a Knudsen number of 0.2. Ren *et al.* [19] numerically examined wall partial slip on a rotating circular cylinder, treating the partial slip as a pattern of alternating shear-free and no-slip boundary conditions. They utilized the gas fraction (*GF*) to represent wall partial slip in this unique setup and explored non-dimensional rotation rates ranging from 0 to 6 at a Reynolds number of 100. Their findings revealed that wall slip exerts varying influences across different flow regimes; the first critical rotation rate decreases with increasing *GF*, whereas the second and third critical rotation rates increase with *GF*. Some prior studies have addressed three-dimensional scenarios. Kouser *et al.*[20] numerically investigated wall slip on a three-dimensional NACA0012 hydrofoil, considering surface heterogeneity through periodic no-slip and shear-free grates arranged perpendicular to the inlet flow direction. The study was conducted over a range of attack angles from 0 to 250 degrees at a fixed Reynolds number of 1000. They observed that bubble separation is delayed for higher gas fractions, minimizing vortex shedding effects and maintaining a two-dimensional flow for higher angles of attack compared to three-dimensional hydrofoils with no-slip boundaries. Mode C with a smaller wavelength was observed at an angle of attack of 150 degrees. You & Moin[21] conducted a numerical study using direct numerical simulations (DNS) and large eddy simulations (LES) for Reynolds numbers of 300 and 3900. They found that hydrophobic treatment on a microscale circular cylinder results in reduced mean drag and root mean square lift coefficient values. Furthermore, the drag reduction in the laminar vortex shedding regime is primarily attributed to the reduction of surface friction, while in the shear layer transition regime, it is due to a delayed separation.

From the above review, it is evident that although the two methods of describing superhydrophobic surfaces differ, they both exhibit similar behaviors. The slip-no-slip description method, while capable of capturing more details of the flow field near the wall, is computationally very demanding. The Navier slip model offers an efficient alternative. However, the relationship between these two descriptions and how they might be converted into one another is still unknown. The objective of this study is to determine the equivalent wall slip length for flow around a super-hydrophobic cylinder using numerical calculations. In this paper, *Re* are set to range from 0.2 to 180. For *Re* beyond 188,[23] the flow undergoes a three-dimensional transition to mode A in the no-slip scenario. To conserve computational resources, the calculations presented in this paper are confined to two dimensions, thus the maximum $Re$ considered is 180. The structure of this article is as follows: Section 2 details the geometric model and the numerical methodology employed. Section 3 presents an analysis of the equivalent slip length under varying parameter conditions, along with a comparative assessment of flow parameters for the two methodologies in question. The final section provides a succinct conclusion.

## 2. Problem Statement and Governing equations

As depicted in figure 1(a), a two-dimensional simulation of flow around a cylinder with diameter $D$ (where the cylinder radius $a = D/2$ ) is conducted in an infinite domain with a uniform flow speed $U_\infty = 1$. The cylinder's center coincides with the origin of the coordinate system. The $x$-axis aligns with the direction of the incoming flow, while the $y$-axis is oriented perpendicular to it. The dimensions of the computational domain are $L_u + L_v$ in length and $H$ in height. The cylinder is positioned $L_u$ away from the inlet boundary and $L_v$ (downstream length) from the outlet boundary. The governing equations have been non-dimensionalized using the following scaling variables: $D$ for the length variable, including the slip length; $U_\infty$ for the velocity vector $\boldsymbol{u}$; $D/U_\infty$ for time $t$; $\rho U_\infty^2$ for pressure $p$; and $U_\infty/D$ for viscosity $\mu$. For the unsteady, two-dimensional, incompressible flow, the non-dimensionalized governing equations are as follows:

$$\nabla \cdot \boldsymbol{u} = 0,$$
$$\frac{\partial \boldsymbol{u}}{\partial t} + \boldsymbol{u} \cdot \nabla \boldsymbol{u} = -\nabla p + \frac{\Delta \boldsymbol{u}}{Re}, \tag{1}$$

where $Re = \rho U_\infty D/\mu$ is Reynolds number.

The alternating arrangement of no-slip and slip conditions on the cylinder's surface is illustrated in figure 1(a). Under the no-slip condition, indicated by red dotted lines in the figure, the boundary is defined with the velocity vector $\mathbf{u}$ set to zero. Conversely, for the slip condition, represented by black solid lines, the boundary is characterized by setting the shear wall friction vector $\mathbf{f}_w = \mu \frac{d\overrightarrow{u_\tau}}{dn} = \left( \mu \frac{du}{dn}, \mu \frac{dv}{dn} \right)$ to zero, where $\mathbf{n}$ denotes the exterior normal vector direction. The slip-no-slip distribution is axisymmetric with respect to the $x$-axis, with the section facing the incoming flow designated as the slip condition. The smallest unit of this alternating slip-no-slip distribution is termed a monomer. The ratio of the slip length $l_1$ to the total length $l = l_1 + l_2$ of the monomer is denoted by $GF$, while $l/a$ represents the ratio of the monomer length $l$ to the cylinder radius $a$. The values for $GF$ and $l/a$ range from 0.1 to 0.99 and from $\pi/80$ to $\pi/5$, respectively. The equivalent slip boundary conditions are established as described in reference:[14]

$$\begin{gathered} \boldsymbol{n} \cdot \boldsymbol{u} = 0, \\ \boldsymbol{n} \times \boldsymbol{u} = Kn \cdot D \cdot \boldsymbol{n} \times (\boldsymbol{S} \cdot \boldsymbol{n}), \end{gathered} \tag{2}$$

where $Kn = \lambda/D$ represents the Knudsen number, with $\lambda$ being the slip length. The other boundary conditions are specified as follows: a constant velocity at the inlet ($\boldsymbol{u} = [1,0]$), a symmetry boundary condition at the upper and lower boundaries, and a zero-pressure condition at the outlet boundary ($p_\infty = 0$).

In this research, the simulated Reynolds numbers ($Re$) span from 0.2 to 180. For low Reynolds numbers, like 1, a substantial computational spatial domain is essential to mitigate the blockage effect on flow parameters.[24] Consequently, we employ an extensive computational spatial region with dimensions set to $H = 250D$, $L_u = 125D$, and $L_v = 375D$, as depicted in figure 1(a).

Numerical simulations were conducted utilizing the commercial software Ansys FLUENT. While comprehensive details of the numerical solution process can be found in previous works,[16,25-27] this section will focus on recapping the key aspects of the methodology. A two-dimensional laminar segregated solver was employed to address the incompressible flow on a collocated grid arrangement. Both unsteady and steady solvers have been utilized in this study. Additionally, the QUICK (Quadratic Upwind Interpolation for Convective Kinematics) scheme is implemented to discretize the convective terms in the momentum equations. The coupling of pressure and velocity is addressed through the use of a coupled solver. The velocity flux at the interface is set to utilize the least squares cell-based method. When $Re$ is less than 60, steady-state calculation is used. The convergence criterion for the mass conservation equation is set such that the residual error must be less than $10^{-10}$. When *Re* is larger than 60, unsteady calculation is used. For unsteady calculation, the time-step adopted is $0.025D/U_\infty$. For Reynolds numbers below 60, a steady-state calculation is employed. The convergence criterion for the mass conservation equation is set to ensure that the residual error is less than $10^{-10}$. For Reynolds numbers exceeding 60, an unsteady calculation is utilized, with a time step of $0.025D/U_\infty$. The time step is sufficiently small to ensure both the stability and accuracy of the numerical calculations.

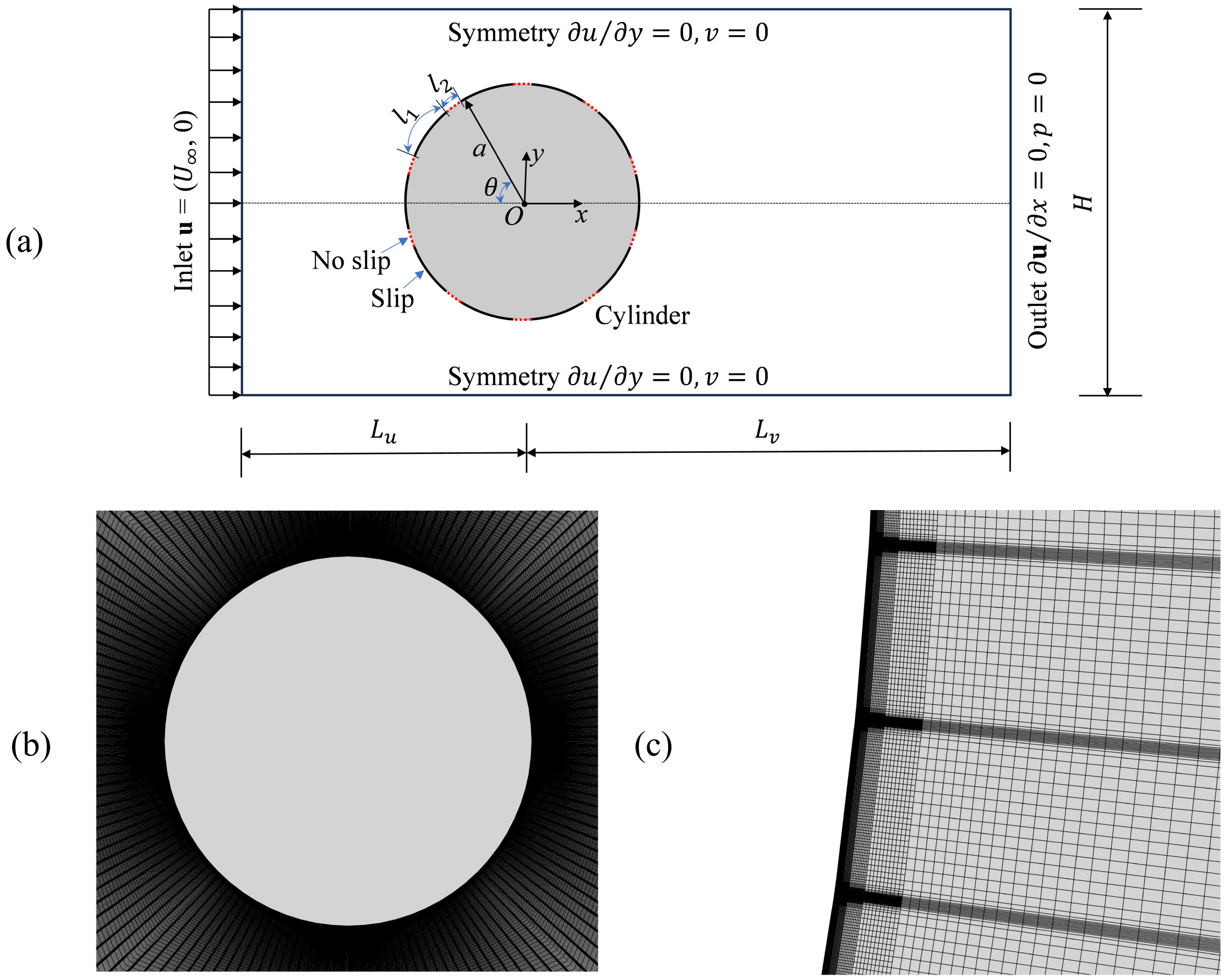

Fig. 1. (a) Schematic of the computational domain. The alternating structure of slip and no-slip conditions encircling the cylinder is delineated in the figure. (b) The mesh around the cylinder. (c) The enlarged view of the mesh near the cylinder surface. The scenario depicted in illustrations (b-c) corresponds to the parameters ($GF$, $l/a$) = (0.95, $\pi/80$).

The commercial grid tool ICEM is employed to generate structured quadrilateral cells with both uniform and non-uniform grid spacing. In the $x$-direction, the downstream region is meticulously equipped with 201 grid points (pertaining to $L_v$) in a non-uniform arrangement, while the upstream region is allocated 51 grid points (pertaining to $L_u$). In particular, the O-type mesh in the vicinity of the cylinder consists of about 30 grid points along the cylinder slip-no-slip monomer (with dense distribution near the discontinuous point between the slip and non-slip boundary condition) and about 201 grid points stretched exponentially along the radial direction, as shown in figure 1(b). For the specific case with parameters ($GF$, $l/a$) set to (0.95, π/80), the cylinder wall is discretized with a total of 4,800 grid points. To minimize the discontinuity effect of the velocity boundary condition, the grid line is purposely located at the junction of the slip and non-slip boundaries, as shown in figure 1(c). To mitigate the discontinuity effects of the velocity boundary conditions, the grid line is intentionally positioned at the transition between the slip and non-slip boundaries, as illustrated in figure 1(c). The radial size of the first cell next to the cylinder surface is set to 0.00005$D$ for ($GF$, $l/a$) = (0.95, π/80). The computational domain encompasses approximately 3,565,000 mesh cells in total for ($GF$, $l/a$) = (0.95, π/80). For other parameter cases, the mesh generation is carried out in a similar manner.

The equivalent slip wall boundary conditions are implemented by introducing wall tangential forces via FLUENT UDF (User-Defined Functions). Equation (2) can be reformulated as follows:

$$\overrightarrow{u_\tau} = \lambda \frac{d\overrightarrow{u_\tau}}{dn}. \tag{3}$$

Moreover,

$$\mathbf{f}_w = \mu \frac{d\overrightarrow{u_\tau}}{dn}. \tag{4}$$

$\mathbf{f}_w$ is shear stress on surface of the cylinder. From equation (3) and equation (4), we can get

$$\mathbf{f}_w = \mu \frac{\overrightarrow{u_\tau}}{\lambda}. \tag{5}$$

Written equation (5) in the form of components in Cartesian coordinate system,

$$f_{wx} = \mu \frac{u}{\lambda} \text{ and } f_{wy} = \mu \frac{v}{\lambda}. \tag{6}$$

The wall shear stress components is loaded on FLUENT software by UDF functions.

A schematic diagram of the angle $\theta$ is shown in figure 1(a), where $\theta$ starts from the front end of the cylinder. The pressure coefficient $C_p$ is defined by the expression:

$$C_p(\theta) = \frac{2(p(\theta)-p_\infty)}{\rho U_\infty^2}, \tag{7}$$

where $p(\theta)$ is the surface pressure at $\theta$ and $p_\infty$ is the pressure at the outlet boundary. The total drag coefficient $C_d$, that is, the sum of the friction drag coefficient and the pressure coefficient, can be written as follows,

$$C_d = \frac{2F_x}{\rho U_\infty^2 D}, \tag{8}$$

where $F_x$ is the total drag force acting on the cylinder. The friction and pressure drag coefficients $C_{dv}$ and $C_{dp}$ are obtained by integrating on the cylindrical surface as follows,

$$C_{dv} = \frac{2\int_0^{2\pi} f_{wx} d\theta}{\rho U_\infty^2}, \tag{9}$$

and

$$C_{dp} = \int_0^{2\pi} C_p \cos\theta d\theta, \tag{10}$$

respectively, where $f_{wx}$ is shear stress component in $x$ direction on surface of the cylinder.

The slip velocity at the wall can be derived using the following formula:

$$u_\tau = u \cdot \sin\theta + v \cdot \cos\theta. \tag{11}$$

For flow around a confined circular cylinder with no-slip boundary, when $Re$ is over than about 6.5,[24] two symmetry recirculation bubbles occurs behind the cylinder, as shown figure 2. When Reynolds number is over than about 46.5,[14] although the flow becomes unsteady, there is still recirculation bubbles behind the circular cylinder in the time-averaged flow field, and the recirculation bubbles still maintains the symmetry up and down. The recirculation bubble length measured with the distance ($L_D \cdot D$) between the end point of the cylinder and the bubble stationary point, which could applied to evaluate the characteristics of wake field. The time-averaged bubble length is recorded to $\overline{L_D}$•$D$.

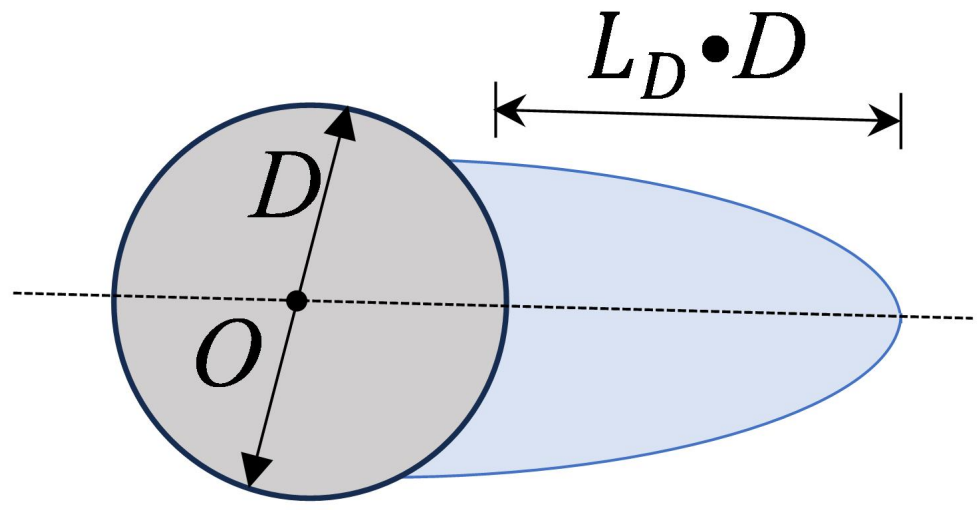

Fig. 2. Schematic diagram of recirculation bubble behind a circular cylinder.

Throughout the simulation, time-averaged flow field computations are performed for more than 10 time periods after the flow has reached its final state. The time-averaged drag coefficient and the root mean square lift coefficient are calculated as follows,

$$\overline{C_d} = \frac{1}{m}\sum_{i=1}^{m} C_d\left(t_i\right), \quad t_i > t_o, \tag{12}$$

$$C_{lrms} = \sqrt{\frac{1}{m-1}\sum_{i=1}^{m}\left[C_l\left(t_i\right)\right]^2}, \quad t_i > t_o, \tag{13}$$

where $t_o$ represents the time instant when the flow reaches statistical stationary state, and $m$ is the total number of statistical moments. In this paper, adding a bar above a variable denotes the time-averaged value of this variable. Another time-averaged parameter ($\overline{C_{dv}}$, $\overline{u_\tau}$, $\overline{L_D}$, $\overline{u_{avg}}$) could be obtained according the above way.

The Strouhal number is introduced to quantify the frequency ($f$) of vortex shedding and defined as follows,

$$St = \frac{fD}{U_\infty}. \tag{14}$$

$f$ is obtained by fast Fourier transform (FFT) of the time series of the lift coefficient when $t >$ $t_o$.

## 3. Results and Discussions

From the definition formula of friction drag coefficient equation (6) and equation (9), we can get

$$\overline{C_{dv}} = 2\pi \frac{\overline{u_{avg}} / U_\infty}{Re \cdot Kn}, \tag{15}$$

where $\overline{u_{avg}}$ is average $x$-velocity on cylindrical wall. If $Kn$ is approach to 0, $\overline{u_{avg}} / U_\infty$ is near zero. From equation (15), we have

$$\overline{u_{avg}} / U_\infty \propto Kn. \tag{16}$$

That is to say, the average flow velocity near the wall is directly proportional to the Knudsen number ($Kn$) for a fixed $Re$.

From equation (15), the equivalent Knudsen number could be calculated by the following equation,

$$Kn = \frac{2\pi}{Re \cdot \overline{C_{dv}}} \cdot \frac{\overline{u_{avg}}}{U_\infty}. \tag{17}$$

Where $\overline{u_{avg}} / U_\infty$ and $\overline{C_{dv}}$ could be obtained by integrating the velocity and friction in the cylinder wall surface.

When Reynolds number ($Re$) does not exceed 1, the flow behavior tends to approximate Stokes flow, which is characterized by $Re$ being close to 0. For the Stokes flow regime, Lauga and Stone,[28] along with Philip,[29,30] have derived several scaling laws for channel flow under certain extreme conditions, such as when the gas fraction ($GF$) approaches 1. Subsequently, we investigate the existence of analogous scaling relationships for the Stokes flow. We set the Reynolds number to 1 for this analysis. As illustrated in figure 3(a), for all values of $l/a$, $\frac{2\pi a}{l} Kn$ across various gas fractions ($GF$), converge onto a single trend, which can be expressed by the following formula:

$$\frac{2\pi a}{l} Kn = f\left(GF\right). \tag{18}$$

That is,

$$Kn = \frac{l}{2\pi a} f\left(GF\right). \tag{19}$$

Where $f$ is a function of $GF$ that is independent of $l/a$. The formula implies for the Stokes flow, $Kn$ is is proportional to $l/a$.

When $GF \geq 0.8$ (which is a natural extension of the scenario where $GF$ approaches 1), $\frac{2\pi a}{l} Kn$ is linearly proportional to $\ln\left(1-GF\right)$, as demonstrated in figure 3(a). This relationship can be formulated as follows:

$$\frac{2\pi a}{l} Kn = A \cdot \ln\left(1-GF\right) + B. \tag{20}$$

That is,

$$f\left(GF\right) = A \cdot \ln\left(1-GF\right) + B. \tag{21}$$

Where $A$ and $B$ are −0.5007 and −0.2169, respectively. Interesting, $A \approx -1/2$, and

$$Kn \approx \frac{l}{4\pi a} \cdot \ln\left(\frac{2/\pi}{1-GF}\right). \tag{22}$$

This formula suggests that an analytical solution to the problem may be within reach, an endeavor that could be pursued in future studies.

Subsequently, we examine the impact of the Reynolds number ($Re$) on the equivalent Knudsen number. Initially, $GF$ is set at 0.95. The Knudsen number is normalized against its value at $Re$ = 1, as illustrated in figure 3(b). When $Re$ does not exceed 2, there is no significant influence of $Re$ on the Knudsen number. However, beyond $Re$ = 2, the Knudsen number decreases with increasing $Re$. This decreasing trend is particularly evident at low values of $l/a$. In our simulations, when $l/a$ is less than or equal to $\pi/40$, the relationship between $\frac{Kn}{Kn_{Re\,=\,1}}$ and ($Re$, $l/a$) can be described by the following formula:

$$\frac{Kn}{Kn_{Re\,=\,1}} = \frac{1}{1+\alpha \cdot Re^{\beta}}, \tag{23}$$

where,

$$\ln(\alpha) = -1.9139 \cdot \ln^2\left(\frac{l}{a}\right) + 0.7 \cdot \ln\left(\frac{l}{a}\right) - 6.0101,$$

$$\beta = 0.2816 \cdot \ln^2\left(\frac{l}{a}\right) - 0.06412 \cdot \ln\left(\frac{l}{a}\right) + 1.0492,$$

where $\pi/40 \leqslant l/a \leqslant \pi/5$. This formula indicates that the *Re* has a diminishing effect on *Kn* when *Re* is low, and as *Re* becomes high, a linear scaling law emerges between ln(*Kn*) and ln(*Re*).

As depicted in figure 3(b), when $l/a=\pi/80$, *Re* exerts no significant influence on *Kn*. With $l/a$ held constant at $\pi/80$, the impact of *GF* is examined, as shown in figure 3(c). It is evident that, for all values of *GF*, at a fixed $l/a=\pi/80$, *Re* has no discernible effect on *Kn*.

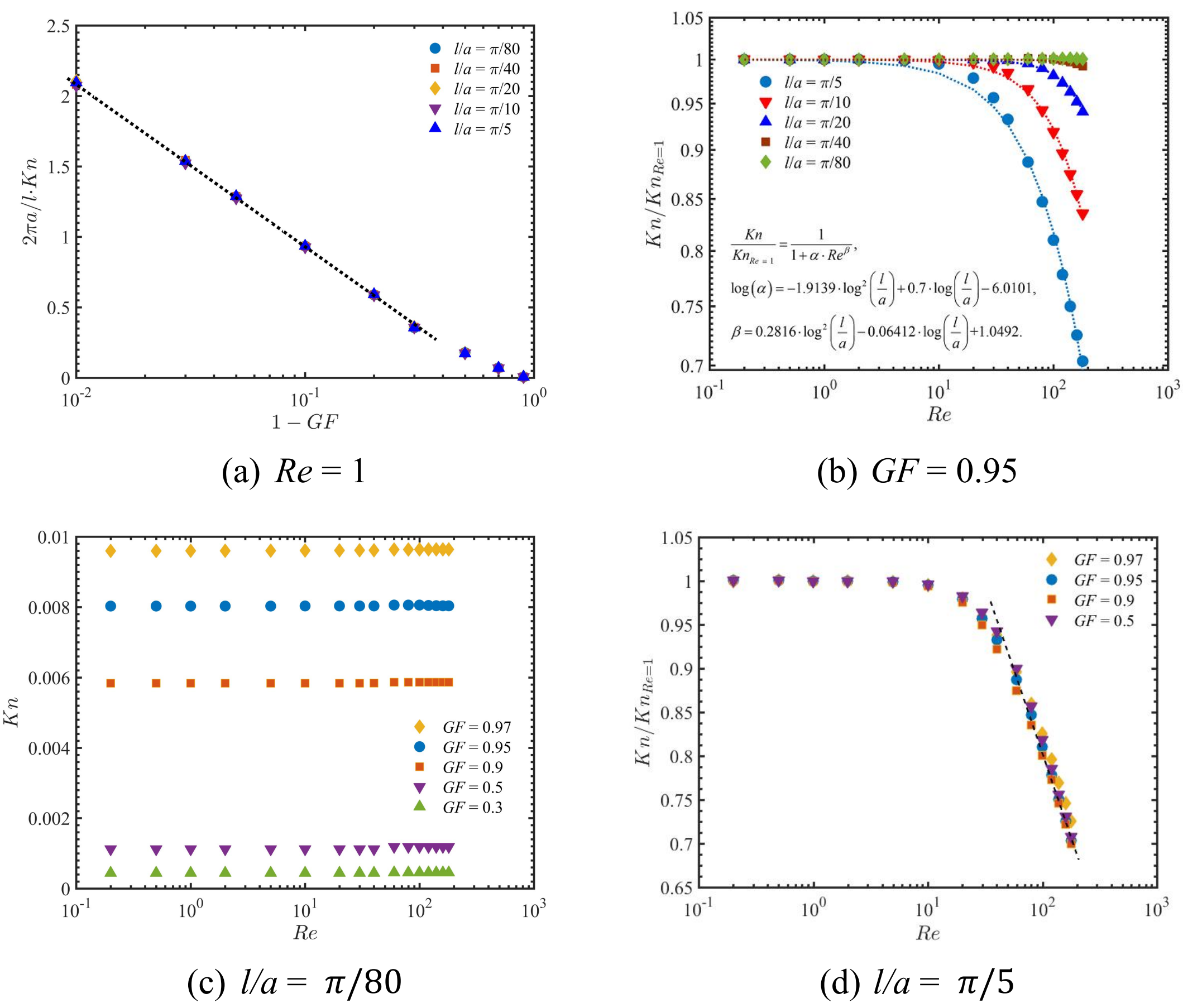

**Fig. 3.** The equivalent slip length. (a) $2\pi a \cdot Kn/l$ with $(1 - GF)$ for a fixed $Re$ = 1. (b) $Kn/Kn_{Re=1}$ with *Re* for a fixed *GF* = 0.95. (c) *Kn* with *Re* for a fixed $l/a = \pi/80$. (d)

$Kn/Kn_{Re=1}$ with $Re$ for a fixed $l/a = \pi/5$.

Subsequently, another $l/a = \pi/5$ is considered. As outlined in equation (23), it is observed that with an increase in $Re$, $Kn/Kn_{Re=1}$ significantly decreases, particularly when $Re$ is high. With $l/a$ fixed at $\pi/5$, the influence of $GF$ is analyzed, as illustrated in figure 3(d). It is clear that, for all calculated values of $GF$, at a constant $l/a = \pi/5$, $GF$ has no significant impact on $Kn/Kn_{Re=1}$ .

Figure 4 illustrates the comparison of tangential velocities along the cylindrical wall between the two approaches. Despite the differing velocity distributions near the cylindrical wall, this is due to the inherent differences in the two boundary description methods. Nevertheless, the flow velocity on the cylinder wall exhibits a consistent general trend: as theta increases, the flow velocity initially rises, peaks at theta = 0.28π, then gradually declines. When theta reaches 0.64π, the velocity drops to zero, and beyond this point, it becomes negative, indicating the presence of recirculation bubbles. The equivalent slip length model serves to mitigate the discontinuities of the cylindrical wall caused by the alternating structure, ensuring that the flow velocity at the wall exhibits a consistent integral behavior. The integral behavior of the flow field will be discussed in the following paragraph.

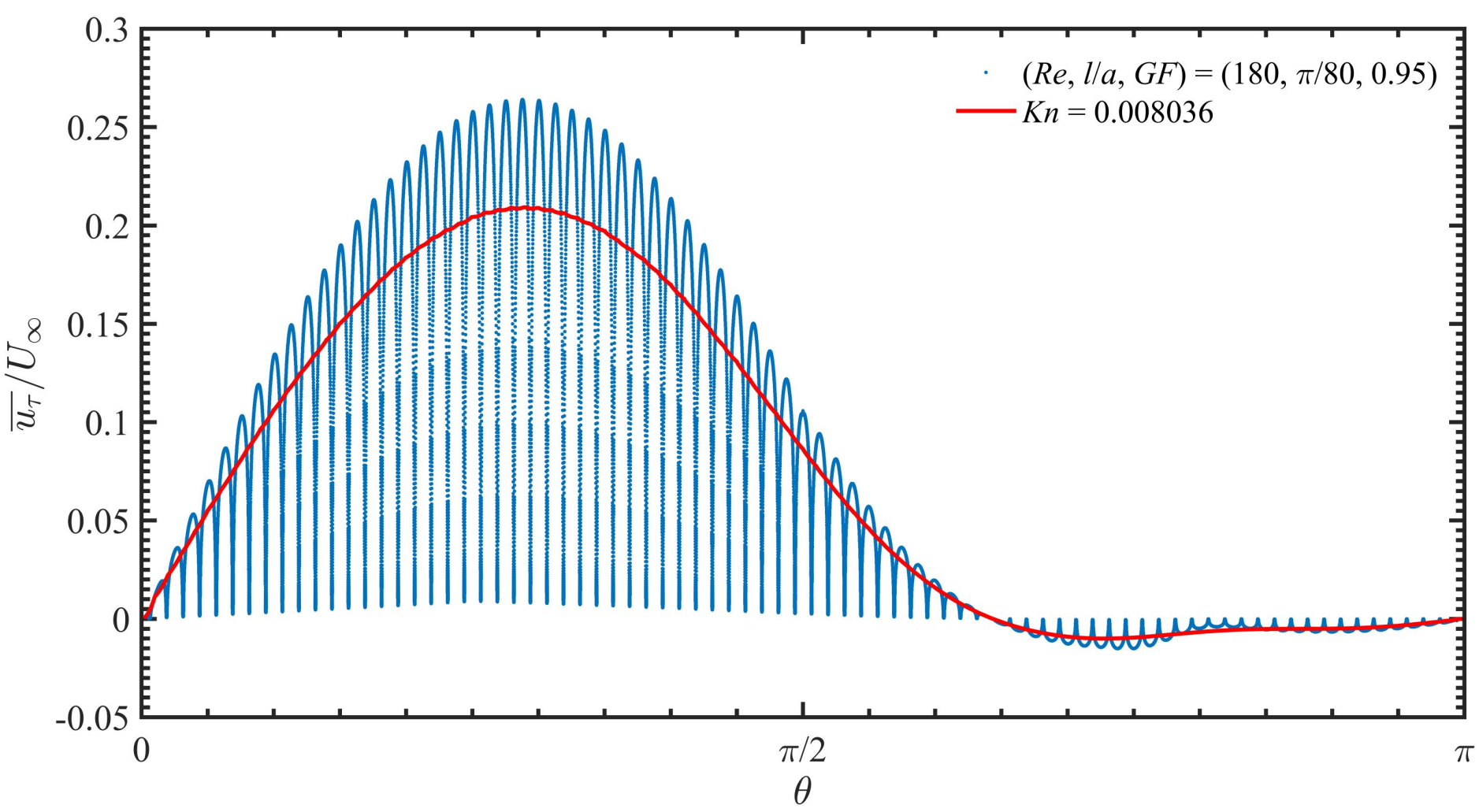

**Fig. 4.** A comparison of the tangential velocity at the cylindrical wall between these two methods.

**Tab. 1.** A meticulous comparison between the two methodologies for the same equivalent slip length.

| ($Re$, $l/a$, $GF$) | (1, $\pi/80$, 0.95) | (60, $\pi/80$, 0.95) | (180, $\pi/80$, 0.95) | (1, $\pi/5$, 0.95) | (180, $\pi/5$, 0.95) | (180, $\pi/5$, 0.5) |
|---|---|---|---|---|---|---|
| $Kn$ | 0.008031 | 0.008057 | 0.008036 | 0.1288 | 0.09063 | 0.01218 |
| $\overline{C_d}$ | 10.275857 | 1.361021 | 1.285675 | 8.247370 | 0.732747 | 1.242254 |
| | 10.226984 | 1.353842 | 1.269856 | 8.277432 | 0.738125 | 1.252020 |
| $\overline{C_{dv}}$ | 4.998136 | 0.408007 | 0.241171 | 3.201874 | 0.133841 | 0.230943 |
| | 4.993860 | 0.406834 | 0.241502 | 3.203988 | 0.137174 | 0.232336 |
| $C_{lrms}$ | - | 0.102036 | 0.422482 | - | 0.140889 | 0.398769 |
| | - | 0.096951 | 0.416197 | - | 0.142946 | 0.389793 |
| $St$ | - | 0.136333 | 0.191388 | - | 0.225606 | 0.194363 |
| | - | 0.135410 | 0.191847 | - | 0.227402 | 0.193424 |
| $\bar{u}$ | 0.006388 | 0.031390 | 0.055518 | 0.065641 | 0.347485 | 0.080587 |
| | 0.006378 | 0.031089 | 0.055827 | 0.065683 | 0.356343 | 0.081294 |
| $\overline{L_D}$ | 0 | 1.895036 | 0.892620 | 0 | 1.099437 | 0.898749 |
| | 0 | 1.921458 | 0.906944 | 0 | 1.040067 | 0.910177 |

**Note:** Uplink represents the results of alternating slip and no-slip conditions, while downlink denotes the calculated outcome of the equivalent slip length.

The flow parameters for the equivalent slip length model, as compared to the slip-no-slip scenario, are detailed in Table 1. The parameters we considered including $\overline{C_d}$, $\overline{C_{dv}}$, $\overline{C_{lrms}}$, $St$, $\bar{u}$ and $\overline{L_D}$. $\overline{C_d}$, $\overline{C_{dv}}$, $\overline{C_{lrms}}$ and $\bar{u}$ can be utilized to delineate the integral behavior of both the flow field or the forces exerted on the cylindrical wall. $\overline{L_D}$ could be employed to assess the time-averaged behavior of the flow field in regions distant from the cylinder wall. $St$ reflects the overall fluctuation frequency of the wake vortex shedding. Our numerical simulations demonstrate that the equivalent slip length model accurately captures the integral behavior of the wall flow field and forces, the time-averaged behavior of the far-field flow field, and the fluctuation frequency of the flow field of the original slip-no-slip model, showing a high degree of correspondence. For example, for ($Re$, $l/a$, $GF$) = (180, $\pi/80$, 0.95), $\overline{C_d}$ for the slip-no-slip model is 1.285675, whereas for the corresponding slip length model, it is 1.269856; $\overline{C_{dv}}$ for the slip-no-slip model is 0.241171, whereas for the corresponding slip length model, it is 0.241502; $C_{lrms}$ for the slip-no-slip model is 0.422482, whereas for the corresponding slip length model, it is 0.416197; *St* for the slip-no-slip model is 0.191388, whereas for the corresponding slip length model, it is 0.191847; $\overline{u}$ for the slip-no-slip model is 0.055518, whereas for the corresponding slip length model, it is 0.055827; $\overline{L_D}$ for the slip-no-slip model is 0.892620, whereas for the corresponding slip length model, it is 0.906944. Hence, the slip length model serves as an effective approximation for the slip-no-slip alternating model, especially when considering macroscopic parameters.

## 4. Conclusion Remarks

In this study, a numerical simulation is conducted to investigate the equivalent wall slip length in the context of flow around a circular cylinder with a super-hydrophobic surface. The super-hydrophobic surface is modeled as an alternating distribution of slip and no-slip conditions along the cylinder's surface. The study takes into account the Reynolds number and two critical dimensionless parameters: the gas fraction (*GF*) and the ratio *l*/*a*. The ranges considered for the Reynolds number, *GF*, and *l*/*a* are from 0.2 to 180, 0.1 to 0.99, and π/80 to π/5, respectively.

A dimensionless number, the Knudsen number (*Kn*), is introduced to measure the ratio between the equivalent slip length and the cylinder's radius. By equating the integral wall friction resistance on the cylinder surface, a quantitative relationship between the equivalent *Kn* and the parameters (*Re*, *GF*, *l*/*a*) is established: (1) For the Stokes flow regime ($Re \to 0$), it has been observed that *Kn* is proportional to the ratio *l*/*a*. Furthermore, when $GF \geq 0.8$, which can be considered a natural extension of the scenario where *GF* approaches 1, $\frac{2\pi a}{l} Kn$ is linearly proportional to $\ln\left(GF\right)$ and $Kn = \frac{l}{4\pi a} \cdot \ln\left(\frac{2/\pi}{1-GF}\right)$. (2) The influence of *Re* on *Kn* is obtained. There is generally an inverse relationship between *Kn* and *Re*, which can be characterized by a specific equation, $\frac{Kn}{Kn_{Re=1}} = \frac{1}{1+\alpha \cdot Re^{\beta}}$, where $\alpha$ and $\beta$ are parameters that are solely dependent on the ratio $l/a$. For a small value of $l/a$, such as $\pi/80$, across all calculated *GF*, *Kn* is found to be almost independent of *Re*. For a relatively high value of $l/a = \pi/5$, across all calculated *GF*, $\frac{Kn}{Kn_{Re=1}}$ is found to be almost independent of *GF*.

It has been observed that the equivalent slip length model effectively smooths out the discontinuities on the cylindrical wall introduced by the alternating structure, ensuring that the flow velocity at the wall maintains a consistent integral behavior. Thus, to the extent that the integral behavior of the wall flow field and forces, the time-averaged behavior of the far-field flow field, and the fluctuation frequency of the flow field, exhibits qualitative and quantitative consistency.

**Acknowledgment**

This research was supported by National Natural Science Foundation of China (Grant Nos.12202102, 12472251, 12072125), The authors are grateful to the support from Guangdong Provincial key Laboratory of Distributed Energy Systems.